\documentclass[letterpaper,journal]{IEEEtran}
\usepackage{amsmath,amssymb,amsfonts,bm,amsthm}
\usepackage{graphicx,booktabs,array,cite}
\usepackage[caption=false,font=footnotesize]{subfig}
\usepackage{microtype,url,flushend}
\usepackage{mathtools,xcolor}
\newif\ifshowchanges
\ifdefined\CleanVersion\showchangesfalse\else\showchangestrue\fi

\usepackage{algorithm,algorithmic}
\usepackage[hidelinks]{hyperref}
\graphicspath{{figures/}}
\hypersetup{pdftitle={Isostable-Based Model Reduction for Optimal Damping of Nonlinear Power System Oscillations},pdfauthor={Kaiyang Huang},pdfsubject={Research manuscript draft},pdfkeywords={isostable, nonlinear model reduction, optimal control, power systems}}

\allowdisplaybreaks
\begin{document}
\title{Isostable-Based Nonlinear Model Reduction for Power System Oscillations}
\author{Kaiyang Huang,~\IEEEmembership{Member,~IEEE, }Dan Wilson, Kai Sun,~\IEEEmembership{Fellow,~IEEE. }
        % <-this % stops a space
\thanks{This work was supported by NSF grants ECCS-2329924 and CMMI-2140527.}
\thanks{K. Huang is with the Department of Electrical Engineering and Computer Science, South Dakota State University, Brookings, SD 57007 USA, and also with the Department of Electrical Engineering and Computer Science, University of Tennessee, Knoxville, TN 37996 USA (e-mail: kaiyang.huang@sdstate.edu).}%
\thanks{D. Wilson and K. Sun are with the Department of Electrical Engineering and Computer Science, University of Tennessee, Knoxville, TN 37996 USA (e-mail: dwilso81@utk.edu,kaisun@utk.edu).}}% <-this % stops a space}

\maketitle
\begin{abstract}
Power systems are often represented by high-dimensional dynamical models, making nonlinear control design computationally expensive and limiting its practical application. Oscillation analysis and damping control therefore commonly rely on linearized models, which do not capture the nonlinear oscillatory behavior induced by large disturbances. This paper proposes an isostable-based model reduction method for nonlinear modal analysis of power systems. The method constructs an invariant manifold associated with a selected oscillatory mode and uses isostable coordinates to describe the mode with only two real state variables. The isostable formulation gives linear autonomous modal dynamics, with nonlinear state reconstruction and input response computed on the modal manifold. An optimal damping control problem is then formulated directly in the reduced coordinates. Case studies on the IEEE 39-bus system demonstrate the accuracy of the proposed model in describing nonlinear oscillations and show improved damping compared with control designed using a linear reduced model.
\end{abstract}

\section{Introduction}\label{sec:intro}

\IEEEPARstart{P}{ower} systems are represented by high-dimensional nonlinear models, making nonlinear analysis and control design computationally demanding. Thus, oscillation analysis and damping control commonly use linearized models. Selective modal analysis reduces their dimension by identifying states associated with the modes of interest \cite{Perez1982}, while coherency-based reduction aggregates generators with similar responses \cite{Chow2013}. For damping design, sensitivities and participation factors guide controller placement and tuning \cite{Pagola1989}, and selected modal models focus the design on a target oscillation \cite{Zhu2018}. However, large disturbances produce nonlinear modal interactions and responses beyond the linear approximation \cite{SanchezGasca2005}. A reduced model that captures these effects is therefore needed.

Normal form analysis introduces higher-order coordinate transformations
to describe nonlinear modal interactions and improve approximations of
power system responses
\cite{Liu2006,Tian2018}. Nonlinear modal decoupling eliminates intermodal
terms to a prescribed order, representing the system as approximately
decoupled nonlinear oscillators \cite{Wang2018}. Related transformations
also define nonlinear participation factors that describe how state
participation changes during a disturbance \cite{Huang2026}. Extending
these methods to higher orders requires successive homological equations
and increasingly many coefficients, while a finite-order expansion need
not remain accurate throughout the basin of attraction. Alternative nonlinear reductions approximate
input-output behavior using empirical covariances \cite{Qi2017} or select
the required model detail according to the disturbance response and modal
participation \cite{Osipov2018,Sajjadi2022}. For control of a selected
oscillation, a reduced model must also relate its nonlinear modal motion
to original states and external inputs.

Invariant-manifold methods provide a geometric basis for this purpose.
Spectral submanifolds extend linear modal subspaces to nonlinear invariant manifolds, with reduced equations determined by the chosen coordinates \cite{Haller2016}. Koopman eigenfunctions define nonlinear
coordinates whose autonomous evolution is linear \cite{Mezic2005}.
Note that Koopman methods have been used for power system coherency identification and stability assessment \cite{Susuki2011,Susuki2014}, and for reduced
prediction in control \cite{Korda2018}. For a stable equilibrium,
principal eigenfunctions define isostable coordinates associated with the
asymptotic decay of individual modes \cite{Mauroy2013}. These coordinates
connect small-signal modal analysis with nonlinear dynamics away from the
equilibrium.

Although autonomous isostable dynamics are linear, physical reconstruction
and input response depend nonlinearly on the modal coordinates.
High-accuracy approximations of these functions support improved prediction
under large-amplitude inputs \cite{Wilson2021}. Related phase-amplitude
and adaptive formulations describe strongly perturbed oscillations using
nonlinear amplitude corrections or families of reference periodic orbits
\cite{Wilson2020,WilsonSun2024}. For a power system mode, the reconstruction,
initial projection, and input response must be combined so that the same
reduced model describes both the physical oscillation and its response to
control.

This paper proposes an integrated isostable-based formulation for nonlinear modal analysis and damping control of power systems. Established isostable and invariant-manifold results are used to connect the physical system, the selected modal motion, and the response to power inputs within a single two-state model. Furthermore, a state-dependent penalty on input-induced forcing of specified omitted modes is incorporated into the reduced optimal-control problem. Thus, the design accounts for their excitation without propagating their dynamic states. Classical-model tests first assess nonlinear reconstruction and forced-response prediction, and detailed-model tests then evaluate the optimized inputs on the full system. A fixed-duration line-fault study further examines whether the control improvement extends beyond one disturbance.

The organization is as follows. Section~\ref{sec:rom} presents the reduction, Section~\ref{sec:control}
formulates the control problem, and Section~\ref{sec:case} reports the
case studies. Section~\ref{sec:conclusion} concludes the paper.
\section{Reduced-Order Modeling}\label{sec:rom}

In this section, a reduced-order model (ROM) is developed for power system oscillations, in which a selected oscillatory mode of the full nonlinear system is represented by two real isostable coordinates. Nonlinear modal projection, physical-state reconstruction, and state-dependent input response are combined within this formulation. The effects of control inputs on the selected and specified omitted modes are then evaluated for the damping control formulation in Section III.

\subsection{Power system model and linear modal analysis}

The dynamical behavior of a power system with external control inputs can be represented by
\begin{equation}\label{eq:full}
    \dot{\mathbf{x}}
    =\mathbf{F}(\mathbf{x})+\mathbf{B}\mathbf{u}(t),
\end{equation}
where $\mathbf{x}\in\mathbb{R}^{n}$ is the state vector,
$\mathbf{F}:\mathbb{R}^{n}\rightarrow\mathbb{R}^{n}$ is the nonlinear vector field,
$\mathbf{u}\in\mathbb{R}^{m}$ is the control input vector, and
$\mathbf{B}\in\mathbb{R}^{n\times m}$ specifies how the inputs act on the system states.
Let $\mathbf{x}_e\in\mathbb{R}^{n}$ be an equilibrium satisfying
$\mathbf{F}(\mathbf{x}_e)=\mathbf{0}_n$.
The state description is taken in independent coordinates, with any common-angle symmetry removed. The equilibrium is assumed to be a hyperbolic sink: the Jacobian $\mathbf{A}=D_{\mathbf{x}}\mathbf{F}(\mathbf{x}_e)\in\mathbb{R}^{n\times n}$ is diagonalizable and all its eigenvalues have strictly negative real parts. All $n$ eigenvalues are counted, including both members of each complex-conjugate pair. To introduce the asymptotic definition below, they are initially ordered by decay rate,
\begin{equation}\label{eq:modeorder}
0>\operatorname{Re}\lambda_1\geq\operatorname{Re}\lambda_2\geq\cdots\geq\operatorname{Re}\lambda_n.
\end{equation}
Thus, $\lambda_1$ has the slowest decay, its conjugate, if present, is placed at index~2. 
For each indexed eigenvalue $\lambda_j\in\mathbb{C}$, let
$\mathbf{v}_j,\mathbf{w}_j\in\mathbb{C}^{n}$ denote its right and left eigenvectors, respectively. These eigenvectors form a biorthogonal basis satisfying
\begin{equation}\label{eq:eig}
    \mathbf{A}\mathbf{v}_j
    =\lambda_j\mathbf{v}_j,
    \mathbf{w}_j^{\mathrm T}\mathbf{A}=\lambda_j\mathbf{w}_j^{\mathrm T},
    \mathbf{w}_j^{\mathrm T}\mathbf{v}_k
    =\delta_{jk},
\end{equation}
where $\delta_{jk}$ is the Kronecker delta, ${\mathrm T}$ denotes the ordinary transpose, and an overbar denotes complex conjugation. Every right eigenvector is assigned unit Euclidean norm in the stated physical coordinates. Its largest angle component fixes the phase to be real and positive, with conjugate vectors chosen consistently. Left eigenvectors are obtained from the inverse right-eigenvector matrix to satisfy \eqref{eq:eig}, without separate unit normalization which is used throughout model construction and control comparisons. For a complex eigenvalue,
$\lambda_j=\alpha_j+\mathrm{i}\beta_j$,
where $\alpha_j<0$ is its decay rate and $\beta_j\ne0$ is its signed angular frequency. Let the state deviation be
$\Delta\mathbf{x}=\mathbf{x}-\mathbf{x}_e\in\mathbb{R}^{n}$
and the linear modal coordinate
$z_j=\mathbf{w}_j^{\mathrm T}\Delta\mathbf{x}\in\mathbb{C}$.
Its evolution under the nonlinear system is
\begin{equation}\label{eq:lincoord}
\begin{aligned}
    \dot z_j
    ={}&\lambda_j z_j
       +\mathbf{w}_j^{\mathrm T}\mathbf{B}\mathbf{u}+\mathbf{w}_j^{\mathrm T}
       \bigl[\mathbf{F}(\mathbf{x})
             -\mathbf{A}\Delta\mathbf{x}\bigr],
\end{aligned}
\end{equation}
The last term prevents independent linear modal evolution. A nonlinear coordinate transformation is therefore required to account for these contributions while preserving the connection to the equilibrium modes.

\subsection{Isostable coordinates and their modal interpretation}
Let $\mathcal{D}\subseteq\mathbb{R}^{n}$ be a forward-invariant
subset containing $\mathbf{x}_e$ within its basin of attraction.
Denote the flow of \eqref{eq:full} by $\boldsymbol{\Phi}_t:\mathcal{D}\rightarrow\mathcal{D}$ when $\mathbf{u}=0$, where $\boldsymbol{\Phi}_t(\mathbf{x})$ is the state reached after time $t$ from $\mathbf{x}$. For the slowest mode, the isostable coordinate function
$\psi_1:\mathcal{D}\rightarrow\mathbb{C}$
is defined by
\begin{equation}\label{eq:limit}
    \psi_1(\mathbf{x})
    =
    \lim_{\tau\rightarrow\infty}
    e^{-\lambda_1\tau}
    \mathbf{w}_1^{\mathrm T}
    \bigl[
        \boldsymbol{\Phi}_{\tau}(\mathbf{x})
        -\mathbf{x}_e
    \bigr].
\end{equation}
The projection describes the slowest modal component as the trajectory approaches equilibrium. Removing its accumulated decay and phase rotation assigns a nonlinear coordinate to the initial state \cite{Mauroy2013,Wilson2021}.
For the other modes, the isostable coordinates $\psi_j:\mathcal{D}\rightarrow\mathbb{C}$ are defined by principal Koopman eigenfunctions \cite{KVALHEIM2021132959}:
\begin{equation}\label{eq:koopman}
\psi_j\bigl(\boldsymbol{\Phi}_t(\mathbf{x})\bigr)=e^{\lambda_jt}\psi_j(\mathbf{x}).
\end{equation}
Here, a principal eigenfunction vanishes at $\mathbf{x}_e$ and has a nonzero gradient there. The functions used below are assumed to be at least twice continuously differentiable, and conjugate eigenvalues are assigned conjugate functions. Their normalization is
\begin{equation}\label{eq:localpsi}
\psi_j(\mathbf{x})=\mathbf{w}_j^{\mathrm T}\Delta\mathbf{x}+O(\|\Delta\mathbf{x}\|_2^2).
\end{equation}
For the slowest coordinate, this normalization is already
fixed by \eqref{eq:limit}, whereas it is imposed explicitly
when the other eigenfunctions are constructed. For an
analytic vector field with the assumed linearization,
a unique local analytic representation in isostable
coordinates is guaranteed under this normalization when
the eigenvalues satisfy
\begin{equation}\label{eq:nonresonance}
    \lambda_j\ne\sum_{k=1}^{n}\ell_k\lambda_k,
    \qquad
    \sum_{k=1}^{n}\ell_k\geq2,
\end{equation}
for all $j$ and multi-indices
$\boldsymbol{\ell}\in\mathbb{N}_0^n$.
For finitely differentiable systems, corresponding
existence and uniqueness results are available under
suitable nonresonance and spectral-spread conditions
\cite{KVALHEIM2021132959}. A full chart additionally requires rank $n$ of the independent real coordinate gradients, guaranteed locally by biorthogonality and continuity but not assumed throughout the basin. For a quadratic nonlinear remainder, $\operatorname{Re}\lambda_j>2a$, where $a=\max_k\operatorname{Re}\lambda_k<0$, is sufficient for the limit in \eqref{eq:limit}. Faster coordinates may require higher-order terminal corrections or direct construction from \eqref{eq:koopman}. Their level sets and autonomous dynamics follow from the eigenfunction identity regardless of this limit.
The isostables associated with mode $j$ are then defined as
the level sets
\begin{equation}\label{eq:isostablelevel}
    \mathcal{S}_{j,r}
    =
    \left\{
        \mathbf{x}\in\mathcal{D}:
        |\psi_j(\mathbf{x})|=r
    \right\},
    \qquad r>0.
\end{equation}
Thus, $\psi_j$ assigns a modal coordinate to a physical state,
whereas $\mathcal{S}_{j,r}$ contains the states sharing
the same coordinate magnitude \cite{Mauroy2013}.
For an oscillatory mode, these states can differ in modal
phase and in the contributions of other modes. These definitions determine how the coordinates evolve.
Let $\psi_j(t) \coloneqq \psi_j(\mathbf{x}(t))\in\mathbb{C}$
denote the coordinate value along a trajectory. 
Differentiating \eqref{eq:koopman} along an unforced trajectory
gives
\begin{equation}\label{eq:autonomouspsi}
    \dot{\psi}_j=\lambda_j\psi_j.
\end{equation}
Unlike a linear approximation of the original system,
\eqref{eq:autonomouspsi} describes the exact evolution
of the nonlinear coordinate throughout its domain of definition when $\mathbf{u}=0$.
Its agreement with linear modal analysis near the equilibrium
follows from \eqref{eq:localpsi}. For an oscillatory mode with
$\lambda_j=\alpha_j+\mathrm{i}\beta_j$, write
$\psi_j=r_j e^{\mathrm{i}\theta_j}$,
where $r_j=|\psi_j|\geq0$ is the modal amplitude and
$\theta_j=\arg\psi_j$ is the modal phase, defined modulo
$2\pi$ when $r_j>0$.
Substitution into \eqref{eq:autonomouspsi} yields
\begin{equation}\label{eq:polarfree}
    \dot r_j=\alpha_j r_j,
    \qquad
    \dot\theta_j=\beta_j.
\end{equation}
All states initially on $\mathcal{S}_{j,r}$ therefore have
modal amplitude $r e^{\alpha_j t}$ after time $t$.
Thus, an isostable level set groups states by the decay of a
selected modal coordinate, rather than by their Euclidean distance from the equilibrium.

Although the modal amplitude decays exponentially and the
modal phase advances uniformly, the corresponding physical
oscillation need not be sinusoidal.
The transformation between physical states and isostable
coordinates is generally nonlinear.
A nonlinear reconstruction is therefore required to describe
the physical oscillation from these modal coordinates.
\subsection{Modal invariant manifold and nonlinear state reconstruction}

To represent a selected oscillation, let $\star$ and $\star^{\mathrm c}$ denote its original eigenvalue indices, with $\lambda_{\star^{\mathrm c}}=\overline{\lambda_{\star}}$ and $\operatorname{Im}\lambda_{\star}>0$. The ordering in \eqref{eq:modeorder} is unchanged. The omitted indices are $\mathcal{J}_{\perp}=\{1,\ldots,n\}\setminus\{\star,\star^{\mathrm c}\}$. Selection may be based on the disturbance response and the chosen pair need not decay slowest.

Where the regular coordinate chart exists, the selected modal manifold is defined by setting all omitted coordinates to zero:

\begin{equation}\label{eq:manifold}
    \mathcal{M}_{\star}
    =
    \left\{
        \mathbf{x}\in\mathcal{D}:
        \psi_j(\mathbf{x})=0,\quad
        j\in\mathcal{J}_{\perp}
    \right\}.
\end{equation}
Each omitted conjugate pair imposes two real constraints and each real coordinate one. Under the chart assumption, the resulting $n-2$ independent constraints define a two-dimensional manifold. Furthermore, \eqref{eq:autonomouspsi} keeps every omitted coordinate at zero, so the manifold is invariant under unforced evolution and its motion is determined by $\psi_{\star}$ alone. To formulate the resulting model in real coordinates, we introduce the reduced state $\mathbf{q}\in\mathbb{R}^{2}$
and the modal matrix
$\mathbf{A}_{\star}\in\mathbb{R}^{2\times2}$ as
\begin{equation}\label{eq:q}
    \mathbf{q}
    =
    \begin{bmatrix}
        \operatorname{Re}\psi_{\star}\\
        \operatorname{Im}\psi_{\star}
    \end{bmatrix},
    \qquad
    \mathbf{A}_{\star}
    =
    \begin{bmatrix}
        \alpha_{\star}&-\beta_{\star}\\
        \beta_{\star}&\alpha_{\star}
    \end{bmatrix}.
\end{equation}
To recover physical states, let $\Omega\subseteq\mathbb{R}^2$ contain the origin and admit a regular parameterization of $\mathcal{M}_{\star}$. The reconstruction $\mathbf{G}:\Omega\rightarrow\mathbb{R}^n$ then defines $\widehat{\mathbf{x}}\in\mathbb{R}^n$ by

\begin{equation}\label{eq:chart}
    \widehat{\mathbf{x}}
    =
    \mathbf{x}_e+\mathbf{G}(\mathbf{q}),
    \qquad
    \mathbf{G}(\mathbf{0}_2)=\mathbf{0}_n.
\end{equation}
By construction, the reconstruction and coordinate
functions satisfy

\begin{equation}\label{eq:chartcoords}
\psi_{\star}(\widehat{\mathbf{x}})=q_1+\mathrm{i}q_2,
\end{equation}
with $\psi_j(\widehat{\mathbf{x}})=0$ for $j\in\mathcal{J}_{\perp}$. On $\mathcal{M}_{\star}$, \eqref{eq:autonomouspsi} gives
$\dot{\mathbf{q}}=\mathbf{A}_{\star}\mathbf{q}$.
Applying the chain rule to \eqref{eq:chart} yields
the invariance equation
\begin{equation}\label{eq:invariance}
    D_{\mathbf{q}}\mathbf{G}(\mathbf{q})
    \mathbf{A}_{\star}\mathbf{q}
    =
    \mathbf{F}\bigl(\mathbf{x}_e+\mathbf{G}(\mathbf{q})\bigr),
\end{equation}
where
$D_{\mathbf{q}}\mathbf{G}(\mathbf{q})
\in\mathbb{R}^{n\times2}$
is the Jacobian of the reconstruction. Thus, the reconstructed velocity equals the original nonlinear velocity on $\mathcal{M}_{\star}$, providing the governing condition for $\mathbf{G}$. The reconstruction can also be computed directly from \eqref{eq:invariance} with the prescribed modal tangent space. This parameterization does not require evaluating every omitted eigenfunction. The zero-coordinate definition in \eqref{eq:manifold} is used where the corresponding regular chart exists. The connection to linear modal analysis follows from
the normalization in \eqref{eq:localpsi}.
Near the equilibrium, the first-order reconstruction
$\mathbf{G}_L:\mathbb{R}^{2}\rightarrow\mathbb{R}^{n}$
is
\begin{equation}\label{eq:linearG}
    \mathbf{G}_L(\mathbf{q})
    =
    2\operatorname{Re}\!\left(
        \mathbf{v}_{\star}(q_1+\mathrm{i}q_2)
    \right).
\end{equation}
The manifold is therefore tangent to the plane spanned by $\operatorname{Re}\mathbf{v}_{\star}$ and $\operatorname{Im}\mathbf{v}_{\star}$. With $\mathbf{G}$ assumed to be $C^3$ near
$\mathbf{q}=\mathbf{0}_2$, its local curvature is
described by the quadratic expansion
\begin{equation}\label{eq:Gexpansion}
    \mathbf{G}(\mathbf{q})
    =\mathbf{G}_L(\mathbf{q})
    +2\operatorname{Re}\!\left(
        \mathbf{g}_{20}\psi_{\star}^{2}
    \right)+\mathbf{g}_{11}|\psi_{\star}|^{2}
    +O(|\psi_{\star}|^{3}),
\end{equation}
where $\mathbf{g}_{20}\in\mathbb{C}^{n}$ and
$\mathbf{g}_{11}\in\mathbb{R}^{n}$ are quadratic
reconstruction coefficients.The $\psi_{\star}^2$ term produces a component at twice the modal phase, whereas $|\psi_{\star}|^2$ shifts the oscillation center with amplitude. Thus, linear coordinate evolution can produce a distorted physical waveform.

This amplitude dependence can also be represented
geometrically using the isostable level sets defined
in \eqref{eq:isostablelevel}.
For a radius $r>0$ whose coordinate circle is contained
in $\Omega$, define
\begin{equation}\label{eq:levels}
    \mathcal{C}_r
    =
    \left\{
        \mathbf{x}_e+\mathbf{G}(\mathbf{q}):
        \mathbf{q}\in\Omega,\quad
        \|\mathbf{q}\|_2=r
    \right\}.
\end{equation}
Each curve lies in $\mathcal{M}_{\star}\cap\mathcal{S}_{\star,r}$, with phase determining position along it. An unforced trajectory crosses successive amplitude levels according to $r(t)=r(0)e^{\alpha_{\star}t}$. While the reconstruction above maps reduced coordinates to
states on the selected modal manifold, a full-system state
$\mathbf{x}\in\mathcal{D}$ need not lie on this manifold.
Its corresponding reduced coordinates are obtained by
evaluating the selected isostable coordinate:
\begin{equation}\label{eq:init}
    \mathbf{q}
    =
    \begin{bmatrix}
        \operatorname{Re}\psi_{\star}(\mathbf{x})\\
        \operatorname{Im}\psi_{\star}(\mathbf{x})
    \end{bmatrix}.
\end{equation}
In particular, the reduced initial condition
$\mathbf{q}(0)=\mathbf{q}_0$ is obtained by evaluating
\eqref{eq:init} at the full-system initial state
$\mathbf{x}=\mathbf{x}_0$.
For $\mathbf{q}_0\in\Omega$, the associated nonlinear
modal projection
$\boldsymbol{\Pi}_{\star}(\mathbf{x}_0)\in\mathcal{M}_{\star}$
is $    \boldsymbol{\Pi}_{\star}(\mathbf{x}_0)
    =
    \mathbf{x}_e+\mathbf{G}(\mathbf{q}_0). $
This projection preserves the selected coordinate and sets the others to zero. Consequently, within the parameterized domain,

\begin{equation}\label{eq:projectionflow}
    \boldsymbol{\Pi}_{\star}
    \bigl(\boldsymbol{\Phi}_t(\mathbf{x}_0)\bigr)
    =
    \boldsymbol{\Phi}_t
    \bigl(\boldsymbol{\Pi}_{\star}(\mathbf{x}_0)\bigr).
\end{equation}
Thus, the reconstructed unforced motion is the nonlinear modal projection of the full trajectory.

\subsection{Isostable response and controlled dynamics}

To extend this unforced modal description to control, the effect of physical inputs on the coordinates is required. For mode $j$, the isostable response $\mathbf{I}_j:\mathcal{D}\rightarrow\mathbb{C}^n$ is defined as the column gradient
 
\begin{equation}\label{eq:Idefinition}
    \mathbf{I}_j(\mathbf{x})
    =\nabla_{\mathbf{x}}\psi_j(\mathbf{x}).
\end{equation}
Its components give the coordinate sensitivity to each physical state. To determine the resulting input response, differentiating \eqref{eq:koopman} gives
\begin{equation}\label{eq:eigenpde}
    \mathbf{I}_j(\mathbf{x})^{\mathrm T}
    \mathbf{F}(\mathbf{x})
    =\lambda_j\psi_j(\mathbf{x}).
\end{equation}
The dynamics in \eqref{eq:full} can be 
expressed in isostable coordinates using
$\dot{\psi}_j
=\mathbf{I}_j(\mathbf{x})^{\mathrm T}\dot{\mathbf{x}}$,
so that substitution of \eqref{eq:eigenpde} gives
\begin{equation}\label{eq:exactpsi}
    \dot{\psi}_j
    =\lambda_j\psi_j
     +\mathbf{I}_j(\mathbf{x})^{\mathrm T}
      \mathbf{B}\mathbf{u}.
\end{equation}
The input sensitivity in these coordinates is determined
by $\mathbf{I}_j(\mathbf{x})$, which varies with the
physical state. To close the selected-mode model, this response is evaluated on the manifold through $\widehat{\mathbf{I}}_j:\Omega\rightarrow\mathbb{C}^n$, defined by

\begin{equation}\label{eq:restrictedI}
    \widehat{\mathbf{I}}_j(\mathbf{q})
    =
    \mathbf{I}_j\bigl(
        \mathbf{x}_e+\mathbf{G}(\mathbf{q})
    \bigr).
\end{equation}
The resulting reduced model is
\begin{equation}\label{eq:complexrom}
\begin{aligned}
    \dot\psi_{\star}
    &=\lambda_{\star}\psi_{\star}
      +\widehat{\mathbf{I}}_{\star}(\mathbf{q})^{\mathrm T}
       \mathbf{B}\mathbf{u}.
\end{aligned}
\end{equation}
The approximation evaluates the input sensitivity at the reconstructed modal state. To distinguish this approximation from the exact coordinate identity, let $\mathbf{q}_{\mathrm F}(t)\in\mathbb{R}^2$ contain the real and imaginary parts of $\psi_{\star}(\mathbf{x}(t))$ along the full controlled trajectory. Its omitted input contribution is
\begin{equation}\label{eq:closureerror}
\varepsilon_{\star}(t)=\bigl[\mathbf{I}_{\star}(\mathbf{x}(t))-\widehat{\mathbf{I}}_{\star}(\mathbf{q}_{\mathrm F}(t))\bigr]^{\mathrm T}\mathbf{B}\mathbf{u}(t).
\end{equation}
Thus, the exact selected-coordinate dynamics equal the reduced right-hand side evaluated at $\mathbf{q}_{\mathrm F}$ plus $\varepsilon_{\star}\in\mathbb{C}$. Control does not change the unforced manifold or its invariance equation, but it generally moves a trajectory away from that manifold. Consequently, the exactness of \eqref{eq:exactpsi} for finite inputs does not make the forced two-state model exact. The resulting prediction error is evaluated using independently projected full-system trajectories in Section~\ref{sec:casecontrol}. For the real coordinate representation introduced in
\eqref{eq:q}, the reduced input matrix
$\mathbf{B}_{\star}:
\Omega\rightarrow\mathbb{R}^{2\times m}$ can be defined as 
\vspace{-2mm}
\begin{equation}\label{eq:reducedB}
    \mathbf{B}_{\star}(\mathbf{q})
    =
    \begin{bmatrix}
        \operatorname{Re}\!\left(
            \widehat{\mathbf{I}}_{\star}(\mathbf{q})^{\mathrm T}
            \mathbf{B}
        \right)\\
        \operatorname{Im}\!\left(
            \widehat{\mathbf{I}}_{\star}(\mathbf{q})^{\mathrm T}
            \mathbf{B}
        \right)
    \end{bmatrix}.
\end{equation}
Then, taking the real and imaginary parts of
\eqref{eq:complexrom} gives
\begin{equation}\label{eq:romphysical}
\begin{aligned}
    \dot{\mathbf{q}}
    &=\mathbf{A}_{\star}\mathbf{q}
      +\mathbf{B}_{\star}(\mathbf{q})\mathbf{u}.
\end{aligned}
\end{equation}
The nonlinear waveform and input response are described by $\mathbf{G}$ and $\mathbf{B}_{\star}$, respectively \cite{Wilson2021}. To recover the linear model, \eqref{eq:localpsi} gives $\mathbf{I}_j(\mathbf{x}_e)=\mathbf{w}_j$ and hence $
    \mathbf{B}_{\star}(\mathbf{0}_2)
    =
    \begin{bmatrix}
        \operatorname{Re}\!\left(
            \mathbf{w}_{\star}^{\mathrm T}\mathbf{B}
        \right)\\
        \operatorname{Im}\!\left(
            \mathbf{w}_{\star}^{\mathrm T}\mathbf{B}
        \right)
    \end{bmatrix}
$. Replacing $\mathbf{G}$ and $\mathbf{B}_{\star}$ by these linear approximations recovers the linear model with the same two dynamic states. For damping design, the nonlinear response can be related directly to the squared modal amplitude $|\psi_{\star}|^2=\mathbf{q}^{\mathrm T}\mathbf{q}$:
 
\begin{equation}\label{eq:amplitudedamping}
\begin{aligned}
    \frac{\mathrm{d}}{\mathrm{d}t}|\psi_{\star}|^2
    ={}&2\alpha_{\star}|\psi_{\star}|^2+2\operatorname{Re}\!\left[
           \bar\psi_{\star}
           \widehat{\mathbf{I}}_{\star}(\mathbf{q})^{\mathrm T}
           \mathbf{B}\mathbf{u}
       \right].
\end{aligned}
\end{equation}
The first term gives natural decay, while a negative second term accelerates the decrease in squared amplitude. Its amplitude and phase dependence determines when an input promotes damping. However, the same input can also excite omitted modes. Since their coordinates vanish on $\mathcal{M}_{\star}$, \eqref{eq:exactpsi} gives

\begin{equation}\label{eq:omittedinput}
    \left.\dot\psi_j\right|_{\mathcal{M}_{\star}}
    =
    \widehat{\mathbf{I}}_j(\mathbf{q})^{\mathrm T}
    \mathbf{B}\mathbf{u},
    \qquad j\in\mathcal{J}_{\perp}.
\end{equation}
These instantaneous forcing terms, together with \eqref{eq:amplitudedamping}, provide the basis for balancing selected-mode damping against control effort and excitation of other modes without adding dynamic states.

\subsection{Computation of the reconstruction and response functions}

The reconstruction and response functions required by
\eqref{eq:romphysical} can be obtained by backward continuation
or by direct expansion in the selected isostable coordinate
and its conjugate \cite{Wilson2021,wilson2026computation}.
Both constructions are based on the invariance equation
\eqref{eq:invariance} and the coordinate dynamics
\eqref{eq:autonomouspsi}.

For backward continuation, a small circle is first specified
in the selected modal coordinate as
$\psi_{\mathrm s}=r_{\mathrm s}e^{\mathrm{i}\theta_{\mathrm s}}$,
where $r_{\mathrm s}>0$ and
$\theta_{\mathrm s}\in[0,2\pi)$.
The quadratic expansion in \eqref{eq:Gexpansion}
is used to initialize the physical state:

\begin{equation}\label{eq:manifoldseed}
\begin{aligned}
    \mathbf{x}_{\mathrm s}
    ={}&\mathbf{x}_e
       +2\operatorname{Re}\!\left(
           \mathbf{v}_{\star}\psi_{\mathrm s}
           +\mathbf{g}_{20}\psi_{\mathrm s}^{2}
        \right)
       +\mathbf{g}_{11}|\psi_{\mathrm s}|^{2}.
\end{aligned}
\end{equation}
The coefficients $\mathbf{g}_{20}$ and $\mathbf{g}_{11}$
are obtained by substituting \eqref{eq:Gexpansion}
into \eqref{eq:invariance} and matching the quadratic terms.
Varying $\theta_{\mathrm s}$ provides initial states around
a small equal-amplitude curve on the local manifold approximation.
The response gradients are initialized using a correction
to $\mathbf{I}_j(\mathbf{x}_e)=\mathbf{w}_j$.
Let
$\mathbf{Q}_j=D_{\mathbf{x}}^2\psi_j(\mathbf{x}_e)
\in\mathbb{C}^{n\times n}$
denote the symmetric coordinate Hessian.
Matching quadratic terms in \eqref{eq:eigenpde} gives

\begin{equation}\label{eq:responsehessian}
\begin{aligned}
    \mathbf{A}^{\mathrm T}\mathbf{Q}_j
    +\mathbf{Q}_j\mathbf{A}
    -\lambda_j\mathbf{Q}_j
    =
    -\sum_{k=1}^{n}w_{j,k}
       D_{\mathbf{x}}^{2}F_k(\mathbf{x}_e).
\end{aligned}
\end{equation}
Here, $w_{j,k}$ and $F_k$ are the $k$th components
of $\mathbf{w}_j$ and $\mathbf{F}$, respectively. The coefficient operator in \eqref{eq:responsehessian}
is nonsingular if
$\lambda_j\ne\lambda_k+\lambda_\ell$
for all full-system eigenvalue pairs.
Its conditioning is checked when these spectral
differences are small.
The response initialization is then

\begin{equation}\label{eq:responseseed}
    \mathbf{I}_j(\mathbf{x}_{\mathrm s})
    \approx
    \mathbf{w}_j
    +\mathbf{Q}_j
     (\mathbf{x}_{\mathrm s}-\mathbf{x}_e).
\end{equation}

To propagate these gradients along the physical trajectories,
the gradient of \eqref{eq:eigenpde} is taken, yielding

\begin{equation}\label{eq:adjointforward}
    \frac{\mathrm{d}\mathbf{I}_j}{\mathrm{d}t}
    =
    \lambda_j\mathbf{I}_j
    -D_{\mathbf{x}}\mathbf{F}(\mathbf{x})^{\mathrm T}
     \mathbf{I}_j,
\end{equation}
where $D_{\mathbf{x}}\mathbf{F}(\mathbf{x})
\in\mathbb{R}^{n\times n}$ is the Jacobian along the trajectory. With backward time $\tau=-t\geq0$ and
$\mathbf{x}_b(0)=\mathbf{x}_{\mathrm s}$,
the state and response equations become

\begin{subequations}\label{eq:back}
\begin{align}
    \frac{\mathrm{d}\mathbf{x}_b}{\mathrm{d}\tau}
        &=-\mathbf{F}(\mathbf{x}_b),\\
    \frac{\mathrm{d}\mathbf{I}_j}{\mathrm{d}\tau}
        &=D_{\mathbf{x}}\mathbf{F}(\mathbf{x}_b)^{\mathrm T}
          \mathbf{I}_j-\lambda_j\mathbf{I}_j,
\end{align}
\end{subequations}
with $\mathbf{I}_j$ evaluated along $\mathbf{x}_b$. The corresponding modal coordinates are assigned using
$\psi_{\star}=\psi_{\mathrm s}e^{-\lambda_{\star}\tau}$,
which gives

\begin{equation}\label{eq:label}
    r=r_{\mathrm s}e^{-\alpha_{\star}\tau},
    \qquad
    \theta=\theta_{\mathrm s}-\beta_{\star}\tau.
\end{equation}
For
$\mathbf{q}=[r\cos\theta,\ r\sin\theta]^{\mathrm T}$,
the integrated states and gradients provide
\begin{equation}\label{eq:continuationdata}
\begin{aligned}
    \mathbf{G}(\mathbf{q})
        &\approx\mathbf{x}_b(\tau)-\mathbf{x}_e,\\
    \widehat{\mathbf{I}}_j(\mathbf{q})
        &\approx\mathbf{I}_j(\mathbf{x}_b(\tau)).
\end{aligned}
\end{equation}
Increasing $\tau$ extends the amplitude, while varying
$\theta_{\mathrm s}$ samples the phase.
The resulting values are assembled using \eqref{eq:label},
and periodic phase interpolation and interpolation between
amplitude levels are used to evaluate both functions.
When omitted modes decay substantially faster,
numerical errors in those directions are amplified
under backward integration.
The constrained continuation procedures in \cite{wilson2026computation}
address this difficulty by enforcing the conditions
associated with the omitted coordinates.

Alternatively, the reconstruction can be computed directly
using the coefficient recursion described in
\cite[Appendix~B]{wilson2021data}.
For this construction, the vector field and the required
coordinate functions are assumed sufficiently smooth
for the retained expansion orders.
With $\psi_{\star}=q_1+\mathrm{i}q_2$,
the reconstruction and response functions are approximated by
\begin{equation}\label{eq:seriesparameterization}
\begin{aligned}
    \mathbf{G}(\mathbf{q})
    &\approx
    \sum_{1\leq k+\ell\leq d_G}
       \mathbf{g}_{k\ell}
       \psi_{\star}^{k}\bar{\psi}_{\star}^{\ell},\\
    \widehat{\mathbf{I}}_j(\mathbf{q})
    &\approx
    \sum_{0\leq k+\ell\leq d_I}
       \mathbf{i}_{j,k\ell}
       \psi_{\star}^{k}\bar{\psi}_{\star}^{\ell},
\end{aligned}
\end{equation}
where $k,\ell\in\mathbb{N}_0$,
$d_G$ and $d_I$ are the truncation degrees,
and $\mathbf{g}_{k\ell},\mathbf{i}_{j,k\ell}
\in\mathbb{C}^{n}$.
The leading coefficients are fixed by
$\mathbf{g}_{10}=\mathbf{v}_{\star}$,
$\mathbf{g}_{01}=\bar{\mathbf{v}}_{\star}$,
and $\mathbf{i}_{j,00}=\mathbf{w}_j$.
Conjugate symmetry,
$\mathbf{g}_{\ell k}=\overline{\mathbf{g}_{k\ell}}$,
is imposed to ensure real reconstructed states.

The reconstruction coefficients are obtained by substituting
\eqref{eq:seriesparameterization} into \eqref{eq:invariance}
and matching equal powers.
Defining
$\sigma_{k\ell}=k\lambda_{\star}
+\ell\bar{\lambda}_{\star}$,
the coefficient equation for $k+\ell\geq2$ is
\begin{equation}\label{eq:Gcoefficients}
    \bigl(\sigma_{k\ell}\mathbf{I}_n-\mathbf{A}\bigr)
    \mathbf{g}_{k\ell}
    =
    \mathbf{r}_{k\ell},
\end{equation}
where $\mathbf{I}_n$ is the identity matrix and
$\mathbf{r}_{k\ell}$ is determined by lower-order
reconstruction coefficients.
The response coefficients are obtained similarly from
\eqref{eq:adjointforward}, with differentiation along
the manifold expressed as
$D_{\mathbf{q}}(\cdot)\mathbf{A}_{\star}\mathbf{q}$.
For $k+\ell\geq1$,
\begin{equation}\label{eq:Icoefficients}
    \bigl[
       \mathbf{A}^{\mathrm T}
       +(\sigma_{k\ell}-\lambda_j)\mathbf{I}_n
    \bigr]\mathbf{i}_{j,k\ell}
    =
    \mathbf{s}_{j,k\ell},
\end{equation}
where $\mathbf{s}_{j,k\ell}$ is determined by the
computed reconstruction and lower-order response coefficients.
Thus, both expansions are obtained by solving linear systems
in increasing total degree. The coefficient matrices and the resulting polynomials are evaluated directly
to obtain $\mathbf{G}$ and $\widehat{\mathbf{I}}_j$. For either construction, the coordinate values required
for initialization can be evaluated using the Hessian
in \eqref{eq:responsehessian}.
The unforced system is integrated forward from $\mathbf{x}$,
and the coordinate is approximated by

\begin{equation}\label{eq:correctedprojection}
\begin{aligned}
    \psi_{j,\tau}(\mathbf{x})
    =e^{-\lambda_j\tau}
    \bigl[
       \mathbf{w}_j^{\mathrm T}\mathbf{d}_{\tau}
       +\tfrac12\mathbf{d}_{\tau}^{\mathrm T}
        \mathbf{Q}_j\mathbf{d}_{\tau}
    \bigr],
\end{aligned}
\end{equation}
where
$\mathbf{d}_\tau
=\boldsymbol{\Phi}_\tau(\mathbf{x})-\mathbf{x}_e$.
With a cubic local coordinate remainder, the
quadratic correction leaves a rescaled error of
$O(e^{(3a-\operatorname{Re}\lambda_j)\tau})$,
which vanishes when $\operatorname{Re}\lambda_j>3a$
and therefore extends the forward-time evaluation
to coordinates satisfying
$3a<\operatorname{Re}\lambda_j<2a$, for which
the quadratic remainder of the linear terminal
approximation may persist under the rescaling. Once the functions are available,
\eqref{eq:romphysical} supplies the two-state dynamics
for control optimization, and \eqref{eq:omittedinput}
supplies the associated forcing of the omitted modes.

\section{Optimal Damping Control Based on the Reduced Model}
\label{sec:control}

Based on the reduced model, a finite-horizon optimal control problem is formulated to damp the selected oscillation subject to physical input limits. To account for the excitation of other modes, a state-dependent quadratic penalty is constructed from their isostable input responses and included in the objective. Thus, input-induced forcing of specified omitted modes is penalized together with control effort, while only the two selected modal states are propagated during optimization.

\subsection{Input constraints and reduced dynamics}

The damping input must satisfy the magnitude limit
$u_{\max,\ell}>0$ of each input channel $u_\ell$.
To express these limits as common unit bounds, the matrix
$\mathbf{U}_{\max}=\operatorname{diag}(u_{\max,1},\ldots,u_{\max,m})
\in\mathbb{R}^{m\times m}$ is introduced. The normalized input
$\boldsymbol{\mu}\in\mathbb{R}^{m}$ is then defined by
\begin{equation}\label{eq:normalizedinput}
    \mathbf{u}(t)=\mathbf{U}_{\max}\boldsymbol{\mu}(t).
    \vspace{-3mm}
\end{equation}
The bounds are then expressed as

$\boldsymbol{\mu}(t)\in\mathcal{U}_{\mathrm N}$, where
\begin{equation}\label{eq:normalizedbounds}
    \mathcal{U}_{\mathrm N}
    =\{\boldsymbol{\mu}\in\mathbb{R}^{m}:
       \|\boldsymbol{\mu}\|_{\infty}\leq1\}.
\end{equation}
Substitution into \eqref{eq:romphysical} introduces the normalized input matrix

$\mathbf{B}_r:\Omega\rightarrow\mathbb{R}^{2\times m}$ as
\vspace{-2mm}
\begin{equation}\label{eq:Br}
    \mathbf{B}_r(\mathbf{q})
    =\mathbf{B}_{\star}(\mathbf{q})\mathbf{U}_{\max}.
\end{equation}
Under an admissible input
$\boldsymbol{\mu}(t)\in\mathcal{U}_{\mathrm N}$, the reduced dynamics
are then written as
\vspace{-2mm}
\begin{equation}\label{eq:rom}
\vspace{-3mm}
    \dot{\mathbf{q}}
    =\mathbf{A}_{\star}\mathbf{q}
     +\mathbf{B}_r(\mathbf{q})\boldsymbol{\mu}.
\end{equation}
\subsection{Penalty on omitted-mode forcing}

An input that damps the selected mode can also excite other coordinates through \eqref{eq:omittedinput}. To account for the omitted modes included in the design, let $\mathcal{K}_{\perp}\subseteq\{j\in\mathcal{J}_{\perp}:\operatorname{Im}\lambda_j\geq0\}$ contain their indices. One member of each included conjugate pair is counted, while an included real coordinate is counted once. The set may contain all omitted modes or a specified subset with computed response functions.

For each coordinate in this set, its normalized input response is obtained by substituting \eqref{eq:normalizedinput} into
\eqref{eq:omittedinput}. The corresponding response
$\mathbf{b}_j:\Omega\rightarrow\mathbb{C}^{1\times m}$ is defined by
\begin{equation}\label{eq:bj}
    \mathbf{b}_j(\mathbf{q})
    =\widehat{\mathbf{I}}_j(\mathbf{q})^{\mathrm T}
     \mathbf{B}\mathbf{U}_{\max},
\end{equation}
and the modal forcing is then described by the scalar function
$g_j:\Omega\times\mathcal{U}_{\mathrm N}\rightarrow\mathbb{C}$, i.e.,
\begin{equation}\label{eq:omittedforcing}
    g_j(\mathbf{q},\boldsymbol{\mu})
    =\mathbf{b}_j(\mathbf{q})\boldsymbol{\mu}.
\end{equation}
At a state on $\mathcal{M}_{\star}$, this quantity is the
instantaneous rate at which the input excites the omitted coordinate.
Consequently, the sum of $|g_j|^2$ over $\mathcal{K}_{\perp}$ measures
the combined forcing of the omitted modes. To include this measure
in the objective as a quadratic input penalty, the matrix
$\mathbf{S}_{\perp}:\Omega\rightarrow\mathbb{R}^{m\times m}$
is defined as
\begin{equation}\label{eq:spill}
    \mathbf{S}_{\perp}(\mathbf{q})
    =\sum_{j\in\mathcal{K}_{\perp}}
     \operatorname{Re}\!\left[
       \mathbf{b}_j(\mathbf{q})^{\mathrm H}
       \mathbf{b}_j(\mathbf{q})\right],
\end{equation}
where $\mathrm H$ denotes the conjugate transpose. The following
identity is then obtained by substituting \eqref{eq:omittedforcing}
into this sum and using the fact that $\boldsymbol{\mu}$ is real:
\begin{equation}\label{eq:spillmeaning}
    \boldsymbol{\mu}^{\mathrm T}
    \mathbf{S}_{\perp}(\mathbf{q})\boldsymbol{\mu}
    =\sum_{j\in\mathcal{K}_{\perp}}
     |g_j(\mathbf{q},\boldsymbol{\mu})|^2.
\end{equation}
Thus, $\mathbf{S}_{\perp}$ is symmetric positive semidefinite, and the forcing penalty is evaluated along the two-state trajectory in \eqref{eq:rom}.
This penalty measures instantaneous input forcing at the reconstructed state. It does not bound the accumulated amplitudes of omitted modes or enforce invariance of $\mathcal{M}_{\star}$ under control. Its role is to discourage inputs that directly excite the specified omitted coordinates while keeping the optimization two-dimensional.

\subsection{Finite-horizon optimal damping control}

The forcing penalty in \eqref{eq:spillmeaning} can now be combined
with the normalized input effort. For this purpose, the input-weighting matrix
$\mathbf{R}:\Omega\rightarrow\mathbb{R}^{m\times m}$ is defined as
\begin{equation}\label{eq:weight}
    \mathbf{R}(\mathbf{q})
    =\rho\mathbf{Id}_m+\kappa\mathbf{S}_{\perp}(\mathbf{q}),
\end{equation}
where $\rho>0$ weights the input effort and $\kappa\geq0$ weights the
omitted-mode forcing. Here, $\mathbf{Id}_d\in\mathbb{R}^{d\times d}$
denotes the identity matrix of order $d$. 
By \eqref{eq:spillmeaning}, $\boldsymbol{\mu}^{\mathrm T}\mathbf{R}\boldsymbol{\mu}$ combines effort and omitted-mode forcing, with $\mathbf{R}$ positive definite throughout $\Omega$. These costs are balanced against $|\psi_{\star}|^2=\|\mathbf{q}\|_2^2$ over a horizon $T>0$. With a symmetric positive definite terminal matrix $\mathbf{P}\in\mathbb{R}^{2\times2}$, the control problem is formulated as

\begin{equation}\label{eq:ocp}
\begin{aligned}
    \min_{\boldsymbol{\mu}(\cdot)}\quad
    \mathcal{J}={}&\mathbf{q}(T)^{\mathrm T}
                   \mathbf{P}\mathbf{q}(T)\\
        &+\int_0^T\!\left[
            \|\mathbf{q}\|_2^2
            +\boldsymbol{\mu}^{\mathrm T}
             \mathbf{R}(\mathbf{q})\boldsymbol{\mu}
          \right]\mathrm{d}t,
\end{aligned}
\end{equation}
subject to \eqref{eq:rom}, $\mathbf{q}(0)=\mathbf{q}_0$,
$\mathbf{q}(t)\in\Omega$, and
$\boldsymbol{\mu}(t)\in\mathcal{U}_{\mathrm N}$ for $0\leq t\leq T$. Time arguments in the integrand are suppressed for compactness. The terminal weight represents the selected-mode amplitude remaining after the input is removed at $T$. Since $\mathbf{A}_{\star}$ is Hurwitz, the unforced tail is quadratic and its matrix satisfies
\begin{equation}\label{eq:care}
\mathbf{A}_{\star}^{\mathrm T}\mathbf{P}+\mathbf{P}\mathbf{A}_{\star}=-\mathbf{Id}_2,\qquad \mathbf{P}=-\frac{1}{2\alpha_{\star}}\mathbf{Id}_2.
\end{equation}
Thus, $\mathbf{q}(T)^{\mathrm T}\mathbf{P}\mathbf{q}(T)$ equals $\int_T^\infty\|e^{\mathbf{A}_{\star}(t-T)}\mathbf{q}(T)\|_2^2\,\mathrm{d}t$. It is an unforced modal-tail cost, not an infinite-horizon feedback value function. For the linear benchmark used below, the equilibrium input matrix is denoted by
\begin{equation}\label{eq:B0}
\mathbf{B}_0=\mathbf{B}_r(\mathbf{0}_2)=\mathbf{B}_{\star}(\mathbf{0}_2)\mathbf{U}_{\max}\in\mathbb{R}^{2\times m}.
\end{equation}
With the objective and terminal weight specified, the necessary
optimality conditions relate the input to the predicted modal response.
The costate $\mathbf{p}\in\mathbb{R}^{2}$ is introduced, and the
Hamiltonian
$\mathcal{H}:\Omega\times\mathbb{R}^{2}\times\mathcal{U}_{\mathrm N}
\rightarrow\mathbb{R}$ is defined as
\begin{equation}\label{eq:ham}
    \mathcal{H}
    =\|\mathbf{q}\|_2^2
     +\boldsymbol{\mu}^{\mathrm T}\mathbf{R}\boldsymbol{\mu}
     +\mathbf{p}^{\mathrm T}
      (\mathbf{A}_{\star}\mathbf{q}+\mathbf{B}_r\boldsymbol{\mu}).
\end{equation}
The arguments of $\mathbf{R}(\mathbf{q})$ and
$\mathbf{B}_r(\mathbf{q})$ are omitted in the following optimality
conditions for compactness. For an optimal trajectory that remains
in the interior of $\Omega$, the costate dynamics are then obtained
from $\dot{\mathbf{p}}=-\nabla_{\mathbf{q}}\mathcal{H}$ as
\begin{equation}\label{eq:costate}
    \dot{\mathbf{p}}
    =-2\mathbf{q}-\mathbf{A}_{\star}^{\mathrm T}\mathbf{p}
     -\nabla_{\mathbf{q}}\!\bigl(
       \boldsymbol{\mu}^{\mathrm T}\mathbf{R}\boldsymbol{\mu}
       +\mathbf{p}^{\mathrm T}\mathbf{B}_r\boldsymbol{\mu}\bigr),
\end{equation}
where the gradient is evaluated with $\mathbf{p}$ and
$\boldsymbol{\mu}$ held fixed. The corresponding boundary condition,
$\mathbf{p}(T)=2\mathbf{P}\mathbf{q}(T)$, is obtained from the
terminal penalty in \eqref{eq:ocp}. Furthermore, the Hamiltonian must be minimized over
$\mathcal{U}_{\mathrm N}$ at each time. For given $\mathbf{q}$ and
$\mathbf{p}$, the terms independent of the input are omitted to obtain
\begin{equation}\label{eq:pointwisecontrol}
    \boldsymbol{\mu}^{\star}
    =\operatorname*{arg\,min}_{\boldsymbol{\mu}\in\mathcal{U}_{\mathrm N}}
      \left[\boldsymbol{\mu}^{\mathrm T}\mathbf{R}\boldsymbol{\mu}
      +\mathbf{p}^{\mathrm T}\mathbf{B}_r\boldsymbol{\mu}\right].
\end{equation}
Since $\mathbf{R}$ is positive definite, this pointwise problem is a
strictly convex quadratic program with a unique minimizer.

This uniqueness concerns the input at fixed $\mathbf{q}$ and $\mathbf{p}$, rather than the complete control trajectory. The state dependence of $\mathbf{B}_r$ and $\mathbf{R}$ generally makes \eqref{eq:ocp} nonconvex. Thus, the Pontryagin conditions above are necessary, but are not established here as sufficient for global optimality, even if solved exactly.
When the input bounds are inactive, the optimal input satisfies the
stationarity condition
$2\mathbf{R}\boldsymbol{\mu}+\mathbf{B}_r^{\mathrm T}\mathbf{p}
=\mathbf{0}_m$, which gives
\begin{equation}\label{eq:unconstrained}
    \boldsymbol{\mu}^{\star}
    =-\tfrac12\mathbf{R}^{-1}\mathbf{B}_r^{\mathrm T}\mathbf{p}.
\end{equation}
If this input violates the bounds, the minimizing value is obtained
from the constrained problem \eqref{eq:pointwisecontrol}.

These relations characterize the input along an optimal trajectory. To compute a control sequence, the finite-horizon objective is discretized and optimized as described next.

\subsection{Numerical solution and linear benchmark}
Divide the horizon into $N_{\mathrm c}$ intervals of length
$\Delta=T/N_{\mathrm c}$, with constant input
$\boldsymbol{\mu}_k\in\mathcal{U}_{\mathrm N}$ on
$[k\Delta,(k+1)\Delta)$. All $mN_{\mathrm c}$ input values are
optimized by direct shooting. Integration over one interval defines the state update $\mathcal{F}_{\Delta}:\Omega\times\mathcal{U}_{\mathrm N}\rightarrow\mathbb{R}^2$ and cost $c_{\Delta}:\Omega\times\mathcal{U}_{\mathrm N}\rightarrow\mathbb{R}$, giving
\begin{equation}\label{eq:discreterom}
\mathbf{q}_{k+1}=\mathcal{F}_{\Delta}(\mathbf{q}_k,\boldsymbol{\mu}_k),
\end{equation}
\begin{equation}\label{eq:discreteobjective}
\mathcal{J}_{\Delta}=\mathbf{q}_{N_{\mathrm c}}^{\mathrm T}
\mathbf{P}\mathbf{q}_{N_{\mathrm c}}
+\sum_{k=0}^{N_{\mathrm c}-1}c_{\Delta}(\mathbf{q}_k,\boldsymbol{\mu}_k).
\end{equation}
The linear problem starts from zero input, and its optimized sequence initializes the nonlinear problem. Thus, \eqref{eq:unconstrained} provides a theoretical characterization; the applied open-loop sequence is obtained by direct shooting. For each feasible candidate, a discrete adjoint
$\mathbf{p}_k\in\mathbb{R}^2$ is propagated backward from
$\mathbf{p}_{N_{\mathrm c}}=2\mathbf{P}\mathbf{q}_{N_{\mathrm c}}$
to obtain the objective gradient with respect to every interval input.
Algorithm~\ref{alg:romcontrol} uses these gradients to update the sequence,
including the state dependence of $\mathbf{B}_r$ and
$\mathbf{S}_{\perp}$. 
A bound-constrained quasi-Newton iteration updates the input sequence using this gradient. Termination is assessed by the projected gradient and objective change; these numerical stopping criteria do not certify global optimality. The computed sequence and its convergence status are therefore reported together.

 Derivatives of $\mathcal{F}_{\Delta}$ and
$c_{\Delta}$ are evaluated at $(\mathbf{q}_k,\boldsymbol{\mu}_k)$.
The accepted input is applied to the full system from $\mathbf{x}_0$;
only the two-state reduced-order model is integrated within the optimization.

\begin{algorithm}[!t]
\caption{Reduced-Order Model-Based Optimal Damping Control}
\label{alg:romcontrol}
\small
\begin{algorithmic}[1]
\ifshowchanges\fi
\REQUIRE $\mathbf{x}_0$; reduced-order model and response data; $\mathbf{U}_{\max}$,
$\rho$, $\kappa$, $T$, $N_{\mathrm c}$; integration tolerances;
convergence tolerance and iteration limit.
\ENSURE Input $\mathbf{u}^{\star}(t)$, convergence status, and domain-check status.
\STATE Obtain $\mathbf{q}_0$ from \eqref{eq:init} and $\mathbf{P}$ from \eqref{eq:care}.
\STATE Initialize the linear problem at zero input; initialize the nonlinear problem using the optimized linear sequence.
\REPEAT
\STATE Integrate \eqref{eq:discreterom} from $\mathbf{q}_0$; evaluate interval costs and derivatives.
\STATE Evaluate \eqref{eq:discreteobjective}; set $\mathbf{p}_{N_{\mathrm c}}=2\mathbf{P}\mathbf{q}_{N_{\mathrm c}}$.
\FOR{$k=N_{\mathrm c}-1,\ldots,0$}
\STATE $\nabla_{\boldsymbol{\mu}_k}\mathcal{J}_{\Delta}
\gets\nabla_{\boldsymbol{\mu}}c_{\Delta}
+(D_{\boldsymbol{\mu}}\mathcal{F}_{\Delta})^{\mathrm T}\mathbf{p}_{k+1}$.
\STATE $\mathbf{p}_k\gets\nabla_{\mathbf{q}}c_{\Delta}
+(D_{\mathbf{q}}\mathcal{F}_{\Delta})^{\mathrm T}\mathbf{p}_{k+1}$.
\ENDFOR
\STATE Check bound-constrained first-order convergence using the gradient.
\IF{not converged and iteration limit not reached}
\STATE Update the sequence with a bound-constrained quasi-Newton step.
\ENDIF
\UNTIL{convergence or iteration limit}
\STATE Record the convergence status of the final input sequence.
\STATE Check whether the corresponding reduced trajectory remains
in $\Omega$ over $[0,T]$; record the domain-check status.
\STATE Set $\mathbf{u}^{\star}(t)
=\mathbf{U}_{\max}\boldsymbol{\mu}_k^{\star}$
for $t\in[k\Delta,(k+1)\Delta)$.
\RETURN $\mathbf{u}^{\star}(t)$, convergence status,
and domain-check status.
\end{algorithmic}
\end{algorithm}

For the linear benchmark, define
$z_{\star,0}^{\mathrm L}=\mathbf{w}_{\star}^{\mathrm T}(\mathbf{x}_0-\mathbf{x}_e)$
and initialize
\begin{equation}\label{eq:linearinit}
\mathbf{q}_{\mathrm L}(0)=
[\operatorname{Re}z_{\star,0}^{\mathrm L},\,
\operatorname{Im}z_{\star,0}^{\mathrm L}]^{\mathrm T}.
\end{equation}
The input responses are fixed at equilibrium:
$\mathbf{b}_{j,0}=\mathbf{w}_j^{\mathrm T}\mathbf{B}\mathbf{U}_{\max}$
and $\mathbf{S}_{\perp,0}=\mathbf{S}_{\perp}(\mathbf{0}_2)$, so that
\begin{equation}\label{eq:linearcoefficients}
\begin{aligned}
\mathbf{R}_{\mathrm L}&=\rho\mathbf{Id}_m+\kappa\mathbf{S}_{\perp,0},\\
\dot{\mathbf{q}}_{\mathrm L}&=\mathbf{A}_{\star}\mathbf{q}_{\mathrm L}
+\mathbf{B}_0\boldsymbol{\mu}_{\mathrm L}.
\end{aligned}
\end{equation}
The same problem \eqref{eq:ocp} is solved with
$\mathbf{q}_0$, $\mathbf{B}_r(\mathbf{q})$, and $\mathbf{R}(\mathbf{q})$
replaced by $\mathbf{q}_{\mathrm L}(0)$, $\mathbf{B}_0$, and
$\mathbf{R}_{\mathrm L}$. Input directions, bounds, horizon, and cost
weights are identical, isolating the effects of nonlinear projection
and state-dependent input response.

\section{Case Studies}\label{sec:case}
The proposed method is evaluated on the IEEE 39-bus system using classical and detailed dynamic models. Nonlinear state reconstruction and forced-response prediction are first examined using the classical model. Damping control is then designed using the reduced model and evaluated on the detailed system with synchronous machines and GFM converters. Comparisons with linear ROM control are performed under the same input constraints and control settings, and different fault locations are considered to assess the consistency of the damping improvement.
\subsection{Nonlinear characterization using the classical model}\label{sec:casecharacterization}

\begin{figure}[!t]
\centering
\subfloat[]{\includegraphics[width=.48\columnwidth]{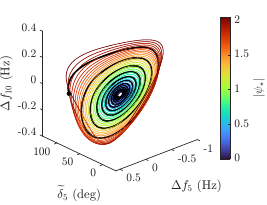}
\label{fig:manifold}}
\hfil
\subfloat[]{\includegraphics[width=.48\columnwidth]{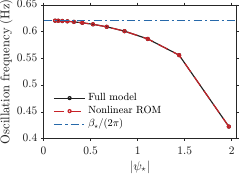}\label{fig:physicalfrequency}}
\caption{Nonlinear characterization of the target mode. (a) Level sets of $\mathcal{M}_{\star}$. The black curve is an unforced full-system trajectory initialized near the level $|\psi_{\star}|=2$. (b) Frequency-amplitude relationship; the blue reference is the small-signal frequency.}
\vspace{-3mm}
\end{figure}

\begin{figure}[!t]
\centering
\subfloat[]{\includegraphics[width=.48\columnwidth]{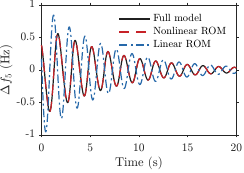}}
\hfil
\subfloat[]{\includegraphics[width=.48\columnwidth]{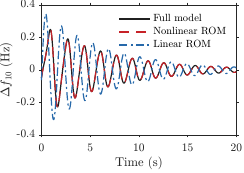}}
\caption{Unforced frequency responses from the same large-amplitude
physical initial state. (a) Generator 5. (b) Generator 10.
The full-system response is compared with the linear and nonlinear
modal reconstructions.}
\label{fig:unforcedfrequencies}
\vspace{-3mm}
\end{figure}

The classical system contains ten generators. Their rotor angles and frequencies are expressed relative to the center of inertia (COI) as
$\widetilde{\delta}_i=\delta_i-\delta_{\mathrm{COI}}$ and
$\Delta f_i=f_i-f_{\mathrm{COI}}$, respectively, for $i=1,\ldots,10$.
The COI values are the corresponding inertia-weighted averages. The lowest-frequency oscillatory mode is selected from the modes with the slowest decay rate and is indexed
by $(\lambda_{\star},\overline{\lambda_{\star}})$. Its small-signal characteristics are given in
Table~\ref{tab:modes}. Generators 10, 5, 9, and 6 have the largest participation factors for this mode, and their responses are used to examine the performance of the proposed model.
\begin{table}[!ht]
\vspace{-3mm}
\centering
\caption{Characteristics of the Selected Oscillation Mode}
\label{tab:modes}
\setlength{\tabcolsep}{5pt}
\begin{tabular}{ccc}
\toprule
Eigenvalues (s$^{-1}$) & Frequency (Hz) & Damping (\%)\\
\midrule
$-0.1500\pm\mathrm{i}3.9010$ & 0.6209 & 3.8423\\
\bottomrule
\end{tabular}
\end{table}
Using the normalization and continuation in Section~\ref{sec:rom}, Fig.~\ref{fig:manifold} shows $\mathcal{M}_{\star}$ in $(\widetilde{\delta}_5,\Delta f_5,\Delta f_{10})$, colored by $|\psi_{\star}|$. The inner curves approach linear modal ellipses, whereas larger amplitudes produce asymmetry and displaced centers.
To relate this geometry to the response, a full-system trajectory starts on the manifold at amplitude 2 and phase 2.75~rad, as shown in black. Both ROMs are initialized by their respective projections of this physical state. In Fig.~\ref{fig:unforcedfrequencies}, the nonlinear reconstruction follows the unequal positive and negative frequency excursions of G5 and G10, whereas the linear model develops substantial amplitude and phase errors.
Over the trajectory, the relative time-integrated $L^2$ error
of the nonlinear reconstruction is 0.0148\% when all ten generator
frequency deviations are compared with the full model. The
corresponding error of the linear reconstruction is 202.7\%.
Thus, the nonlinear geometry of $\mathbf{G}$ is important even
though only two dynamic variables are used and their autonomous
evolution is linear.
The waveform distortion is further examined through its oscillation frequency, calculated as the reciprocal interval between successive upward zero crossings of $\Delta f_5$. Each value is paired with $|\psi_{\star}|$ at the interval midpoint. In Fig.~\ref{fig:physicalfrequency}, the frequency rises from approximately 0.423 to 0.6209~Hz as the amplitude decreases. The nonlinear ROM captures this dependence, whereas the linear model predicts $\beta_{\star}/(2\pi)$. Thus, the nonlinear reconstruction relates the uniform isostable phase evolution in \eqref{eq:polarfree} to a changing physical oscillation period.

To assess input-response prediction, all three models are independently driven from equilibrium by $u_1(t)=a_{\mathrm{in}}\sin(2\pi f_{\mathrm{in}}t)$, where $a_{\mathrm{in}}$ is in p.u./s and $f_{\mathrm{in}}$ in Hz. The input direction is the imaginary part of the selected eigenvector's speed components, scaled to unit maximum magnitude, with other channels zero.

\begin{figure}[!t]
\centering
\subfloat[]{\includegraphics[width=.48\columnwidth]{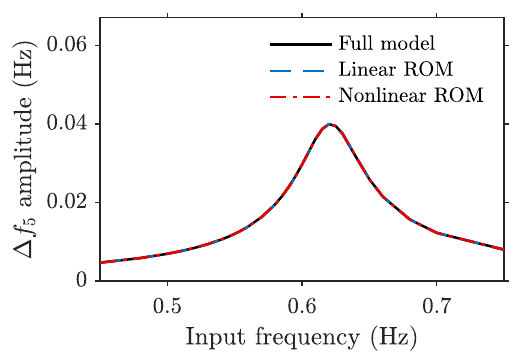}}
\hfil
\subfloat[]{\includegraphics[width=.48\columnwidth]{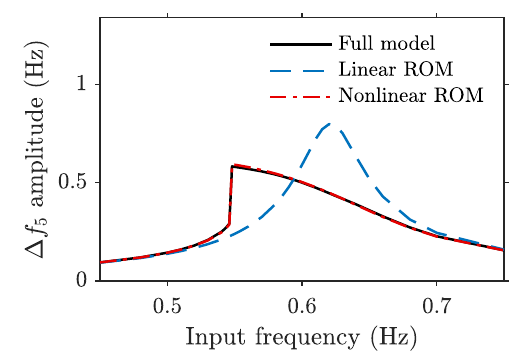}}
\caption{Steady-state fundamental amplitude of $\Delta f_5$ under the same
sinusoidal input in the full system and both reduced-order models.
(a) $a_{\mathrm{in}}=0.0002$~p.u./s. (b) $a_{\mathrm{in}}=0.004$~p.u./s.
Each sampled forcing frequency is simulated independently from equilibrium.}
\label{fig:forcedresponse}
\vspace{-3mm}
\end{figure}

Fig.~\ref{fig:forcedresponse} compares the steady-state fundamental
amplitude of $\Delta f_5$ over forcing frequencies from 0.45 to
0.75~Hz. Each frequency is simulated independently
from equilibrium, and the amplitude is evaluated over five complete
periods after settling. At $a_{\mathrm{in}}=0.0002$~p.u./s, the curves nearly coincide
and peak at the sampled frequency 0.620~Hz. At $a_{\mathrm{in}}=0.004$~p.u./s, the
full-system sampled maximum shifts to 0.548~Hz, with amplitude
0.5834~Hz. The nonlinear reduced-order model predicts 0.5939~Hz at the same frequency,
including the sharp change nearby. The linear reduced-order model instead peaks at
0.620~Hz with amplitude 0.7988~Hz. At the full-system response maximum, the nonlinear amplitude error is 1.80\%.

\begin{figure}[!t]
\centering
\subfloat[]{\includegraphics[width=.48\columnwidth]{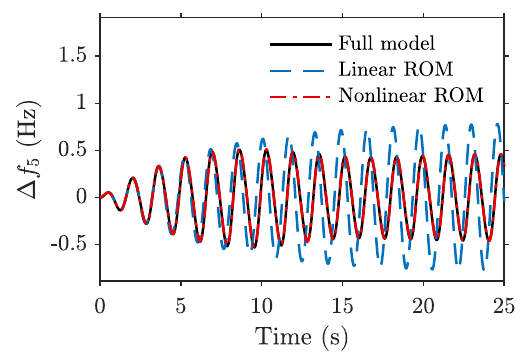}}
\hfil
\subfloat[]{\includegraphics[width=.48\columnwidth]{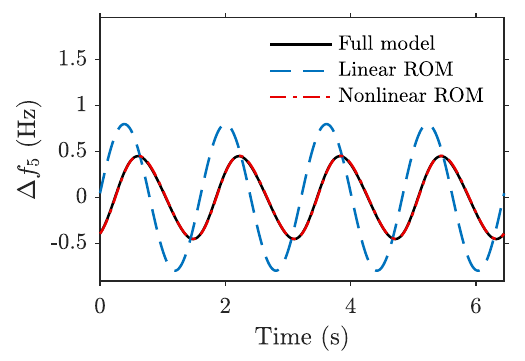}}
\caption{Frequency response of generator 5 under the common sinusoidal
input with $a_{\mathrm{in}}=0.004$~p.u./s and $f_{\mathrm{in}}=0.620$~Hz.
(a) The first 25~s from equilibrium. (b) Four steady-state periods, with
time measured from the start of the displayed window.}
\label{fig:forcedwaveforms}
\vspace{-3mm}
\end{figure}

Fig.~\ref{fig:forcedwaveforms} shows the waveforms at
$a_{\mathrm{in}}=0.004$~p.u./s and $f_{\mathrm{in}}=0.620$~Hz. The nonlinear reduced-order model
follows the initial buildup and subsequent modulation, whereas the
linear prediction develops substantial amplitude and phase errors.
The relative time-integrated $L^2$ errors of $\Delta f_5$ over the
first 25~s are 2.26\% and 119.17\% for the nonlinear and linear reduced-order models.
Over the four steady-state periods, the full-system fundamental amplitude
is 0.4470~Hz, compared with 0.4492~Hz for the nonlinear reduced-order model and
0.7988~Hz for the linear reduced-order model. Their absolute fundamental-phase errors
are $1.11^{\circ}$ and $56.86^{\circ}$, respectively. The relative
waveform errors are 2.16\% and 149.42\%. 

These tests establish nonlinear prediction accuracy. The next subsection examines whether this modeling advantage improves damping control in the detailed system.

\subsection{Optimal damping control using the detailed model}
\label{sec:casecontrol}

The control study uses the modified IEEE 39-bus system with detailed models of synchronous generators and grid-forming converters (GFMs). G1--G3, G8, and G10 are represented by sixth-order synchronous-machine models with excitation and turbine-governor dynamics; G4--G7 and G9 use the GFM models in \cite{Huang2026}. The system has 74 independent states after fixing the common-angle reference.

The 0.5533-Hz pair, $-0.3063\pm\mathrm{i}3.4768$~s$^{-1}$, is selected because it makes the largest contribution to the fault-induced frequency response. For this system, the slowest decay rate is
$a=-0.13784~\mathrm{s}^{-1}$ and the selected
pair has $\alpha_\star=-0.30627~\mathrm{s}^{-1}$,
so $3a<\alpha_\star<2a$ and the quadratic
terminal correction in
\eqref{eq:correctedprojection} is used to obtain
a decaying terminal remainder.

For the detailed model, the reconstruction and isostable
response functions are computed directly from
\eqref{eq:Gcoefficients} and \eqref{eq:Icoefficients}.
The expansions in \eqref{eq:seriesparameterization}
are truncated at $d_G=30$ and $d_I=28$, respectively.
The minimum spectral separations of the reconstruction
and response coefficient equations are
$0.0432$ and $0.1249~\mathrm{s}^{-1}$, respectively.
At the initial reduced state and sampled points along
the optimized trajectory, relative residuals below $1\%$
are required for \eqref{eq:invariance} and the
selected-coordinate identity \eqref{eq:eigenpde}.
Consistency between the real coordinate gradients
and reconstruction derivatives is also checked against
the identity matrix, with a Frobenius-norm tolerance
of $0.02$.

\begin{figure}[!t]
\centering
\subfloat[]{\includegraphics[width=.48\columnwidth]{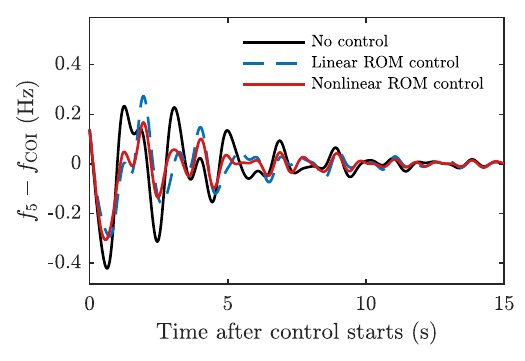}}
\hfil
\subfloat[]{\includegraphics[width=.48\columnwidth]{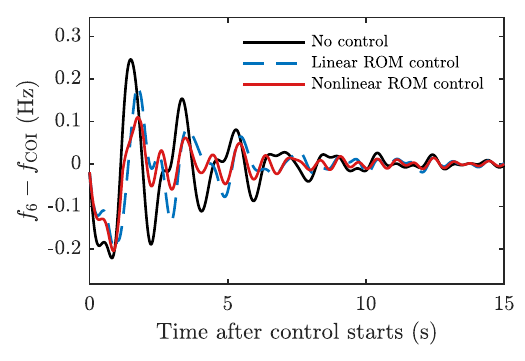}}\\
\subfloat[]{\includegraphics[width=.48\columnwidth]{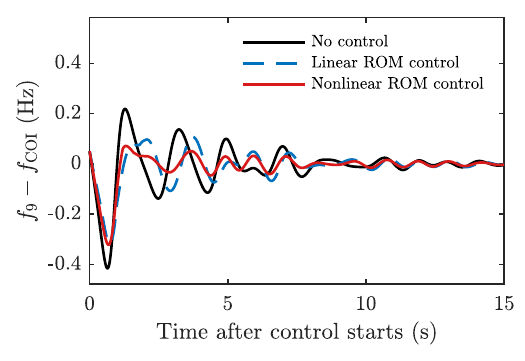}}
\hfil
\subfloat[]{\includegraphics[width=.48\columnwidth]{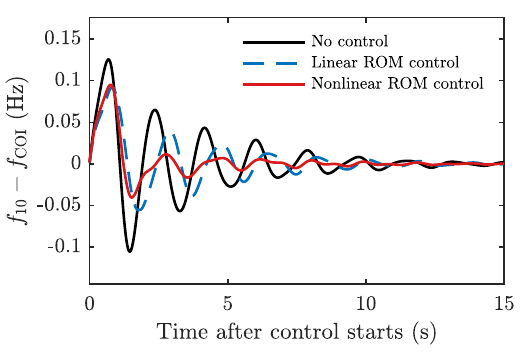}}
\caption{COI-relative frequency responses of the detailed nonlinear
model. (a) G5. (b) G6. (c) G9. (d) G10.}
\label{fig:detailedcontrolfrequency}
\vspace{-2mm}
\end{figure}

\begin{figure}[!t]
\centering
\subfloat[]{\includegraphics[width=.48\columnwidth]{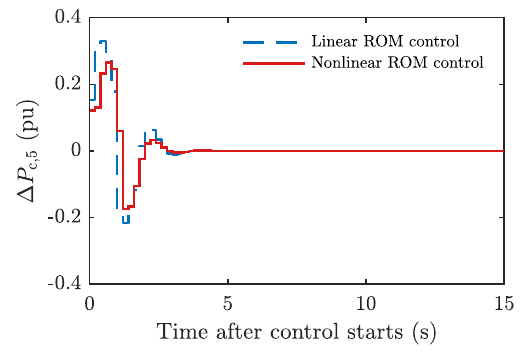}}
\hfil
\subfloat[]{\includegraphics[width=.48\columnwidth]{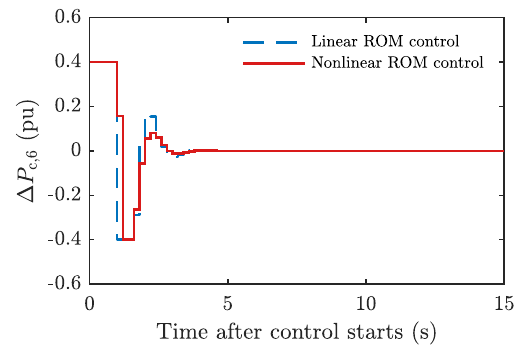}}\\
\subfloat[]{\includegraphics[width=.48\columnwidth]{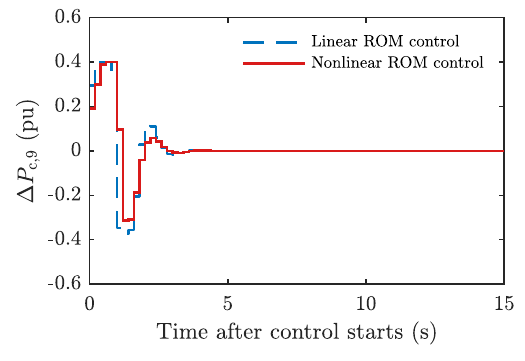}}
\hfil
\subfloat[]{\includegraphics[width=.48\columnwidth]{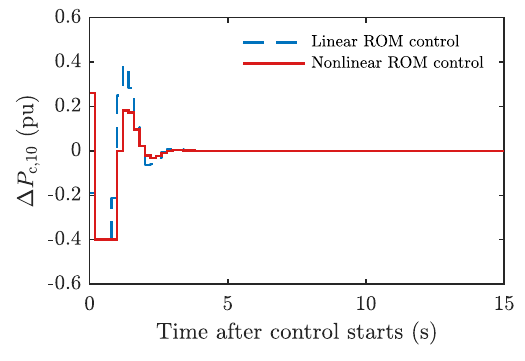}}
\caption{Optimized supplementary active-power commands in per unit.
(a) G5. (b) G6. (c) G9. (d) G10.}
\label{fig:detailedcontrolinput}
\vspace{-2mm}
\end{figure}

A representative control test considers a temporary three-phase fault at bus~3,
which is cleared after eight cycles. The system is allowed to evolve for
0.5~s after clearing before the damping input is activated, and the time
origin in the following control plots is placed at this activation
instant. The same physical state is used to initialize every controlled
and uncontrolled full-system simulation. Its selected isostable
amplitude is $|\psi_{\star}|=0.6912$. The nonlinear input is obtained from \eqref{eq:ocp} using
Algorithm~\ref{alg:romcontrol}. For comparison, a second input is
optimized using the linear single-mode model formed from
$\mathbf{G}_L$ and $\mathbf{B}_{\star}(\mathbf{0}_2)$. Both optimizations use
the same physical initial state, input channels, limits, horizon, and
cost weights. Six supplementary active-power channels are placed at
G4--G7, G9, and G10. The five GFM commands enter their active-power
dynamics, whereas the G10 command enters the synchronous-generator
speed equation as supplementary power. Each channel satisfies
$|u_\ell|\leq0.4$~p.u. The 15-s horizon is divided into 75 intervals, and the normalized effort and
omitted-mode forcing weights in \eqref{eq:weight} are $\rho=0.004$ and
$\kappa=1$, respectively. The omitted-mode penalty includes the four computed response functions at 0.8498, 1.0145, 1.4227, and 1.5128~Hz. Thus, both input trajectories are optimized
by propagating only the two real states in $\mathbf{q}$ and are then
applied in open loop to the same detailed nonlinear model. Fig.~\ref{fig:detailedcontrolfrequency} compares representative
full-system frequency responses. The linear-reduced-order model input reduces the
initial oscillation, but a slowly decaying component remains over much
of the control horizon. The nonlinear-reduced-order model input produces a similar
initial peak and then decreases the subsequent oscillations more
rapidly. This behavior is observed for both GFM and synchronous-machine
frequencies, indicating that the improvement is not confined to the actuated units.
The corresponding active-power commands in Fig.~\ref{fig:detailedcontrolinput} show changes in timing and magnitude. The nonlinear design accounts for $\mathbf{B}_r(\mathbf{q})$ along the predicted trajectory, whereas the linear design uses its equilibrium value.
The full-system damping performance is quantified by
$J_\psi=\int_0^T|\psi_{\star}(\mathbf{x}(t))|^2\,\mathrm{d}t$ and
$J_f=\int_0^T\sum_{i=1}^{10}\Delta f_i(t)^2\,\mathrm{d}t$.
Furthermore, $E_P=\int_0^T\|\mathbf{u}(t)\|_2^2\,\mathrm{d}t$ measures
the physical input effort, and $t_{5\%}$ is the first time after which
$|\psi_{\star}|$ remains below 5\% of its value at control activation. To
distinguish the later response from the initial peak, the system-wide
frequency RMS over the final 5~s is also reported. These quantities are
evaluated from the trajectories of the detailed nonlinear model and are
listed in Table~\ref{tab:detailedperformance}.

\begin{table}[!t]
\centering
\caption{Damping Performance Evaluated on the Detailed Nonlinear Model}
\label{tab:detailedperformance}
\setlength{\tabcolsep}{2.7pt}
\begin{tabular}{lrrr}
\toprule
Index & No control & Linear ROM & Nonlinear ROM\\
\midrule
$J_\psi$ & 0.7799 & 0.4713 & 0.3273\\
$J_f$ (Hz$^2$\,s) & 0.7209 & 0.4162 & 0.3089\\
Peak $|\Delta f_i|$ (Hz) & 0.4208 & 0.3172 & 0.3215\\
Final-5-s frequency RMS (Hz) & 0.00830 & 0.00785 & 0.00662\\
$|\psi_{\star}(\mathbf{x}(T))|$ & 0.006989 & 0.004657 & 0.001470\\
$t_{5\%}$ (s) & 9.8 & 8.5 & 4.7\\
$E_P$ (p.u.$^2$\,s) & 0 & 1.2911 & 1.0100\\
\bottomrule
\end{tabular}
\vspace{-2mm}
\end{table}

Relative to linear ROM control, the nonlinear design reduces $J_f$ by 25.78\%, $J_\psi$ by 30.55\%, and physical effort by 21.77\%. The 5\% amplitude threshold is reached at 4.7~s rather than 8.5~s. Although the initial peak is similar, the subsequent oscillation decays faster, consistently with the integral damping objective in \eqref{eq:ocp}.

To assess the controlled-model approximation in \eqref{eq:closureerror}, the full-system coordinate is independently evaluated using \eqref{eq:correctedprojection}. For the reduced prediction $\widehat\psi_{\star}(t)\in\mathbb{C}$, the relative error is
\begin{equation}\label{eq:controlpredictionerror}
e_\psi=\left[\frac{\int_0^T|\widehat\psi_{\star}(t)-\psi_{\star}(\mathbf{x}(t))|^2\,\mathrm{d}t}{\int_0^T|\psi_{\star}(\mathbf{x}(t))|^2\,\mathrm{d}t}\right]^{1/2}.
\end{equation}
Under the same nonlinear input, $e_\psi=26.16\%$ from the physical fault state and 3.63\% from its modal projection $\boldsymbol{\Pi}_{\star}(\mathbf{x}_0)$. The latter removes initial omitted-mode content while allowing control-induced departure from the manifold. The 50/65-s coordinate estimates differ by only $0.00045\%$ in the same norm. Thus, modal prediction accuracy is assessed separately from the full-system damping benefit.

To examine performance beyond the bus-3 example, all 34 transmission lines are subjected to eight-cycle midpoint three-phase faults without line tripping; the 12 transformer branches are excluded. The operating point, inputs, activation delay, horizon, limits, and weights are unchanged across faults. Of these cases, 29 recover without control and admit coordinate initialization. The selected 0.5533-Hz pair is dominant in all 29 responses after clearing, so the same ROM is used.

Both controls recover in 28 cases, providing complete 15-s paired comparisons. For line 17--27, only nonlinear control recovers. The other five faults, on lines 25--26, 26--27, 26--28, 26--29, and 28--29, fail without control and are not assigned a ROM-control comparison. Incomplete responses are therefore excluded from integral percentage comparisons.

\begin{figure}[!t]
\centering
\includegraphics[width=1\columnwidth]{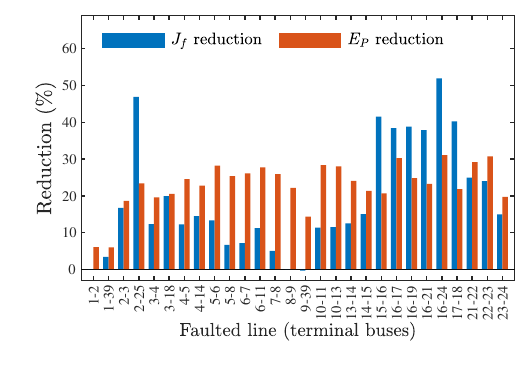}
\caption{Full-system frequency and effort reductions relative to linear ROM control for the 28 paired line faults. Every fault lasts eight cycles and no line is tripped. }
\label{fig:linecensus}
\end{figure}

For each paired case, the reduction of an achieved index $J$ is calculated as $100(J_{\mathrm L}-J_{\mathrm{NL}})/J_{\mathrm L}$. Fig.~\ref{fig:linecensus} shows these reductions for $J_f$ and $E_P$, and Table~\ref{tab:linestatistics} gives the distribution over the 28 cases. Positive values indicate an advantage for the nonlinear design. Frequency reductions within $\pm0.1$ percentage points are classified as ties. The nonlinear design improves $J_f$ in 25 cases, is nearly unchanged in two, and is slightly worse in one. Its control effort is lower in all 28 cases. The mean and median frequency reductions are 19.02\% and 13.93\%, while the corresponding effort reductions are 23.04\% and 23.72\%.

\begin{table}[!t]
\centering
\caption{Nonlinear-Control Improvement Over 28 Paired Line Faults}
\label{tab:linestatistics}
\setlength{\tabcolsep}{5pt}
\begin{tabular}{lrr}
\toprule
Statistic & $J_f$ reduction (\%) & $E_P$ reduction (\%)\\
\midrule
Mean & 19.02 & 23.04\\
Median & 13.93 & 23.72\\
25th percentile & 10.23 & 20.66\\
75th percentile & 28.19 & 27.84\\
Minimum & $-0.31$ & 6.02\\
Maximum & 51.92 & 31.08\\
\bottomrule
\end{tabular}
\end{table}

The largest frequency reduction is 51.92\% for line 16--24. The smallest is a 0.31\% increase for line 9--39, whose effort still decreases by 14.34\%. Furthermore, seven line faults exceed the representative bus-3 improvement. Thus, the nonlinear advantage extends across fault locations in the detailed system.

\section{Conclusion}\label{sec:conclusion}

An isostable-based formulation has been developed for nonlinear modal analysis and damping control of power systems. The selected oscillation is represented by two real dynamic states, while nonlinear reconstruction and input-response functions relate its motion to physical states and power inputs. Furthermore, a penalty on the forcing of specified omitted modes is included in the reduced control objective without adding dynamic states. The results have demonstrated accurate prediction of waveform distortion and amplitude-dependent frequency beyond the linear approximation. Furthermore, this nonlinear representation improves the applied control, and nonlinear modal reduction supports improved damping design across multiple disturbance locations while restricting the optimization to two dynamic states.

\section*{Acknowledgments}
During manuscript preparation, ChatGPT (OpenAI) was used to assist with checking the mathematical derivations in Sections II–III and with language editing. Responsibility for the final content is retained by the authors.
\bibliographystyle{IEEEtran}
\bibliography{references}
\end{document}